\documentclass{emulateapj}
\usepackage{graphics}
\usepackage{multirow}
\usepackage{ifpdf}

\shorttitle{Moat flow in the vicinity of sunspots}
\shortauthors{Vargas Dom\'inguez et al.}

\def\farcs{\hbox{$.\!\!^{\prime\prime}$}}

\begin{document}

\title{Moat flow in the vicinity of sunspots for various penumbral configurations}

\author{S. Vargas Dom\'inguez\altaffilmark{1,2}, L. Rouppe van der Voort\altaffilmark{3,4}, J.A. Bonet\altaffilmark{1}, V. Mart\'inez Pillet\altaffilmark{1}, M. Van Noort\altaffilmark{5}, Y. Katsukawa\altaffilmark{6}}
\altaffiltext{1}{Instituto de Astrof\'isica de Canarias, 38205 La Laguna, Tenerife, Spain; svargas@iac.es, jab@iac.es, vmp@iac.es}
\altaffiltext{2}{Department of Astrophysics, University of La Laguna, 38200, La Laguna, Tenerife, Spain.}
\altaffiltext{3}{Institute of Theoretical Astrophysics, University of Oslo, P.O. Box 1029 Blindern, N-0315 Oslo, Norway; rouppe@astro.uio.no}
\altaffiltext{4}{Center of Mathematics for Applications, University of Oslo, P.O. Box 1053 Blindern, N-0316 Oslo, Norway.}
\altaffiltext{5}{Institute for Solar Physics of the Royal Swedish Academy of Sciences, AlbaNova University Center, SE-10691 Stockolm, Sweden; noort@astro.su.se}
\altaffiltext{6}{National Astronomical Observatory of Japan, 2-21-1 Osawa, Mitaka, Tokyo 181-0033, Japan; yukio.katsukawa@nao.ac.jp}

\clearpage

\begin{abstract}
 High-resolution time  series of sunspots have been obtained with the Swedish 1m Solar
 Telescope between 2003 and 2006 at different locations on the solar disc. Proper motions in seven different active regions have been studied. 
 The analysis has been done by applying local  correlation tracking to every series of sunspots,  each of them more than 40 minutes long. 
  The sunspots' shapes include a different variety of penumbral
  configurations. We report on a systematic behaviour of the large-scale
  outflows surrounding the sunspots, commonly known as moat flows,
  that are essentially present only when preceded by a penumbra not
  tangential but perpendicular to the sunspot border.
  We present one case for which this rule appears not to be
  confirmed. We speculate that the magnetic neutral line, which is
  located in the vicinity of the anomalous region, might be responsible for blocking the
  outflow.
   These new results confirm the systematic and strong relation between
  the moat flows and the existence of penumbrae. A comparative statistical study between moats and standard granulation is also performed.
  
\end{abstract}
\keywords{Sun: sunspots -- Sun: granulation -- Sun: photosphere}

\section{Introduction}

The Sun's strong magnetic behaviour is clearly revealed through its most prominent
visible manifestation, the sunspots. 
The newest generation of solar
telescopes and the lastest restoration techniques have greatly
increased the spatial resolution of the images and therefore the
variety of details and tiny structures in and around
sunspots that can be resolved.
Nevertheless, it is still not well understood how these structures
are formed, evolve and affect the photosphere surrounding
them. 
Many approaches have been done in this direction such as the
identification of small magnetic elements called "Moving Magnetic
Features" (MMF) travelling radially outwards while immersed in an
anullar cell around the sunspot of strong radial outflows known as a
"moat" \citep{sheeley1972,harvey1973}. For a recent overview of the
literature on MMFs see the introduction of \citet{hagenaar2005}. The
sunspot moat thus defines an organized horizontal flow pattern which
ends quite abruptly at a distance that can be comparable to
supergranular sizes or even larger.
\cite{sobotka2007} study the characteristics and temporal evolution of moats at two heights in the atmosphere for a large sample of sunspots with different sizes, shapes and evolutive stage. A similar work studying the properties of moats was previously performed by \cite{brickhouse1988}. The averaged moat flow velocity reported ranges from $\sim$0.4 to $\sim$1 km\,s$^{-1}$.

New findings \citep{sainz2005} have shown that the penumbral filaments extend beyond the
sunspot boundary entering the region dominated by the moat flow where
the MMF activity is detected. Thus, the temporal average of magnetograms has unveiled the existence of moat filaments: horizontal, filamentary structures 
coming from the penumbra and reaching the photosperic network as an
extension of penumbral filaments. Moreover, some MMFs have been
found \citep{sainz2005,ravindra2006} starting just inside the sunspot boundary in its way 
out from the sunspot. \cite{kubo2007} have established a relationship between the vertical components (spines) of the magnetic field in the so-called uncombed structure of the penumbra and MMFs observed in moat regions.

In a recent letter, \citet{vargas2007} studied the moat flow in a
complex active region. They found that no outflow was detected in the
granulation next to umbral cores that lack penumbrae. 
Outflows were only
found in the granulation regions adjacent to the penumbrae in the
direction following the penumbral filaments. Granulation regions
located next to penumbral sides parallel to the direction of the
filaments show no moat flow.

The aim of the present work is to extend the study by  \citet{vargas2007} 
to a larger sample of cases, that is to establish whether 
this moat-penumbra relation is systematically found in other 
active regions and how the granular convective pattern
surrounding sunspots behaves.
In doing so, we have used several high
quality time series observed at high spatial resolution. A total of 7
different sunspot series have been processed and analyzed. 
The sample includes sunspots with different penumbral
configurations, varying from well-developed penumbrae to 
rudimentary penumbral morphologies.
The paper describes first in \S~\ref{sec:obs} how the observations were
performed and the restoration technique used to correct for
atmospheric aberrations. Once the time series are ready to analyze,
\S~\ref{sec:analysis} presents some calculations and statistics of proper motions of structures 
surrounding the sample sunspots. We finally discuss the results and future work in  \S~\ref{sec:dis}.

\section{Observations and data processing}
\label{sec:obs}

The observations were obtained with the \mbox{Swedish 1-m Solar Telescope}
\citep[SST,][]{SST} on La Palma between 2003 and 2006.
All observations benefited from the use of the SST adaptive optics
system \citep{SSTAO} that minimized the degrading effects of seeing. 
Image post-processing techniques were applied to increase the
homogeneity in the quality of the time series and to enhance the image
quality over the whole field-of-view (FOV). 
For the 2003 data set we applied Multi-Frame Blind Deconvolution
(MFBD) using the implementation developed by \citet{lofdahl02MFBD}.
The MFBD code was succeeded by the Multi-Object Multi-Frame Blind
Deconvolution code (MOMFBD, \citet{MOMFBD}) which
employs multiple objects and phase-diversity.
MOMFBD was applied to the data sets after 2003. 
For all the time series we present in this paper, the seeing
conditions were generally very good and sometimes excellent. 
A large fraction of the restored images in the different time series
approach the diffraction limit of the telescope.

After image restoration, we applied standard techniques to the time
series including correction for the diurnal field rotation, rigid
alignment and de-stretching to correct for seeing-induced image
warping. A p-modes filter was also applied to the series (threshold phase velocity 4 ~km\,s$^{-1}$). 

Table~\ref{table1} presents details for the different active region
targets, Table~\ref{table2} gives some details on the restored time
series. 
Below we provide more detailed information on the different data sets. 

\begin{table*}[]
\caption{Sunspots Sample}
\centering
\begin{tabular}{cccccc}
\hline
Name & Active region & Observing date & Heliographic coordinates & $\mu$ \\
     &      NOAA     &&& \\
\hline
S1 & 10440 & 2003, Aug 18 & (S10, W1) & 0.96 \\
S2 & 10608 & 2004, May 10 & (S5, E57) & 0.55 \\
S3 & 10662 & 2004, Aug 20 & (N13, E7) & 0.99 \\
S4 & 10662 & 2004, Aug 21 & (N11, W5) & 0.99 \\
S5 & 10789 & 2005, Jul 13 & (N17, W32) & 0.88 \\
S6 & 10813 & 2005, Oct 04 & (S7, E37) & 0.78 \\
S7 & 10893 & 2006, Jun 10 & (N1, E17) & 0.95 \\
\hline
\end{tabular}
\label{table1}
\end{table*}

\subsection{S1: AR440, 22-Aug-2003}
An interference filter centered at the continuum at 436.4~nm (FWHM
1.1~nm) was used. In the remainder, this filter will be referred to as
``G-cont'' filter.
Every 25~s, the 3 highest contrast images were selected and stored to
disk. 
These 3 images were used for MFBD processing. This sunspot is shown in Figure~\ref{F:2}.

\subsection{S2: AR608, 10-May-2004}
An interference filter centered at the G-band at 430.5~nm (FWHM
1.2~nm) was used. In the remainder, this filter will be referred to as
``G-band'' filter. 
This observation involved 2 cameras set-up as a phase-diversity pair:
one in focus and one slightly out of focus. 
20 exposures from each camera, acquired over the course of 7.5
seconds, were used for MOMFBD processing.
During processing of the time series, a few bad quality images,
suffering from too much blurring, were dropped.
The exposure time was 11~ms. This sunspot is shown in Figure~\ref{F:3}.

\subsection{S3: AR662, 20-Aug-2004}
This time series was observed with the G-cont filter. 
The MOMFBD processing was done using 10 exposures (acquired during 7s)
from a phase-diversity pair of cameras. 
For the central part of the FOV, centered on the sunspot, the MOMFBD
processing was extended to involve a G-band phase-diversity pair of
cameras. 
After MOMFBD restoration, the best quality image from 3 subsequent
images was selected for further processing of the time series.
The exposure time was 11.3~ms. This sunspot is shown in Figure~\ref{F:9}.

\subsection{S4: AR662, 21-Aug-2004}
This time series was processed in a similar manner as S3. After MOMFBD
restorations, the worst 14\% of the images were dropped from the time series. This sunspot is shown in Figure~\ref{F:4}.

\subsection{S5: AR789, 13-Jul-2005}
For this time series, MFBD processing was applied to 18 subsequent
G-band images, acquired during 10~s. 
Images that were too blurred and when the AO system was not actively
compensating, were dropped from the MFBD sets. This sunspot is shown in Figure~\ref{F:5}.

\subsection{S6: AR813, 4-Oct-2005}
This time series was observed with an interference filter centered on
the spectral region between Ca~K and Ca~H at 395.4~nm (FWHM 1.0~nm).
For the MOMFBD processing, 20 exposures (acquired during 8.3~s) from a
phase-diversity pair of cameras were used.
In addition, simultaneous exposures from two other cameras were used
for the restoration: one equipped with a narrow-band filter centered
in the Ca~H wing at 396.5~nm and one equipped with a narrow-band
filter centered on the Ca~H core (396.8~nm).
The exposure time was 11~ms. This sunspot is shown in Figure~\ref{F:6}.

\subsection{S7: AR893, 10-Jun-2006}
This time series was observed with an interference filter centered at
630.2~nm (FWHM 0.8~nm). This is the pre-filter for the Lockheed SOUP
filter. 
MOMFBD processing was applied to 400 exposures obtained during 19~s. 
The restorations included images from a phase-diversity pair of
cameras with the wide band filter, and narrow band SOUP exposures in
the Fe~I~630.2~nm spectral line. This sunspot is shown in Figure~\ref{F:7}.

\begin{table}[]
\caption{Restored sunspots time series}
\centering
\begin{tabular}{cccc}
\hline
Name & N. images & Cadence (s) & Duration (min:s) \\
\hline
S1 & 128 & 24.7 & 52:47 \\
S2 & 376 & 7.5 & 47:10 \\
S3 & 144 & 19.8 & 47:20\\
S4 & 344 & 8.0 & 45:57\\
S5 & 240 & 10.1 & 40:12 \\
S6 & 556 & 8.7 & 80:3 \\
S7 & 124 & 19.7 & 40:26\\
\hline
\end{tabular}
\label{table2}
\end{table}

\section{Data analysis}
\label{sec:analysis}

The high quality, stability and long duration of the restored sunspot series, enable us to follow the
dynamics of the plasma around the sunspots. We have computed
proper-motion velocity fields (horizontal velocities) employing the
local correlation tracking (LCT) technique (\citet{november1988}  using the implementation of 
\citet{molowny1994}). 
A Gaussian tracking window of FWHM $1\farcs0$ suitable for tracking
mainly granules has been used for all the series. The method produces a sequence of map-pairs each one describing, within a short time interval (tens of milliseconds), the horizontal velocity components $-x$ and $-y$ along the FOV. 
These maps of components have been averaged over 5 and 10 min periods, and also over the respective total duration of every sunspot series (more than 40 min in all cases). From the averaged maps of velocity components, the distribution of velocity magnitudes over the FOV (flow map) has been easily derived.

It is well documented in the literature that LCT-techniques in general produce some systematic errors in the determination of displacements (see Figure~13 in \cite{november1988}). Their significance depends on several factors like the width of the tracking window in relation to the size of the elements used as motion tracers, and the presence of large intensity gradients in the images. Also the nature of the interpolation algorithm to fix the position of the correlation maximum with sub-pixel accuracy may play a role. The LCT-method typically underestimate the velocity fields. Tests performed by the authors of the LCT-code we are employing in this work lead to conclude that in the worst cases this underestimation can amount to \mbox{20-30~\%} \citep{yi1992, molownyphd1994}. Even so, this drawback does not change the main conclusions of the present paper since we are not interested in fixing absolute values of velocities but rather in the detection of large-scale regularly organized flows around the sunspots in comparison with those typically found in a  normal granulation field. In both cases we use the solar granules as the tracers of motion and the same size for the tracking window so that in case of some bias it will affect similarly to both velocity fields.

The sample of sunspots studied in this paper includes a variety of
heliocentric positions on the solar disc as shown in
Table~\ref{table1}  (with $\mu=\cos \theta$, where $\theta$ is the
heliocentric angle).  Away from solar disc center,  the measured proper
motions are in fact projections of the real horizontal velocities in the
sunspot plane onto the plane perpendicular to the line-of-sight (LOS).
Thus, to evaluate the real horizontal velocities we have de-projected the velocities obtained from the LCT technique to a plane tangent to the solar surface at the center of the sunspot. 

\begin{figure}
\centering
\begin{tabular}{l}
\includegraphics[width=.9\linewidth]{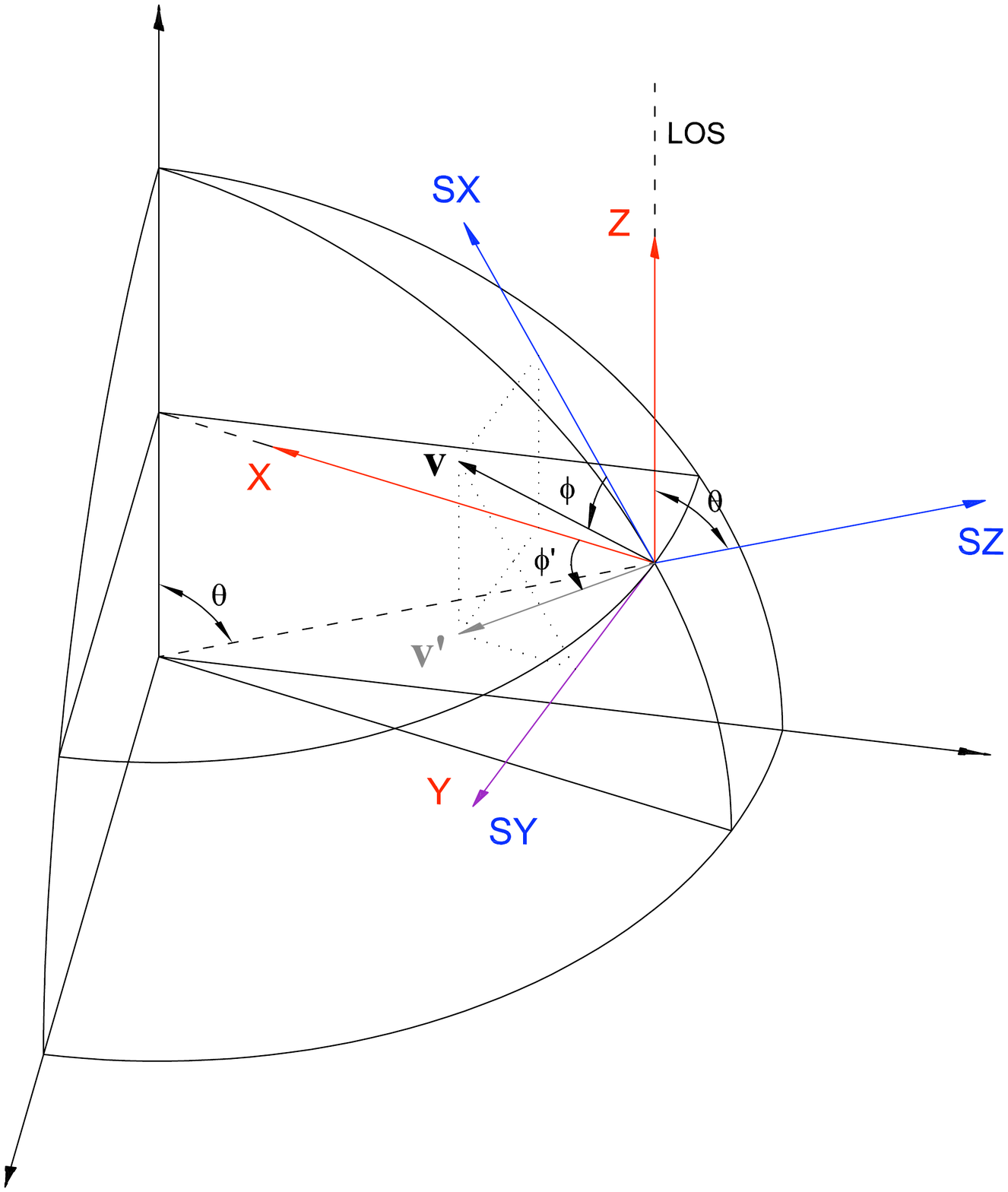}
\end{tabular}
\caption[]{Sketch showing two orthogonal coordinate systems for the projection analysis of the horizontal velocities: 1) The Sunspot System (SX,SY,SZ) with the Z-axis perpendicular to the solar surface at the sunspot location, and 2) the Observing System (X,Y,Z), where the observations are described, with the Z-axis coinciding with the line-of-sight.}
\label{F:1}
\end{figure}

In Figure~\ref{F:1} we consider the orthogonal coordinate system SX,SY,SZ (the Sunspot System, SS) with the SZ-axis
perpendicular to the plane tangent to the solar surface at the sunspot
location and the SX-axis tangent to the meridian of the solar sphere
at this location. The figure also shows the coordinate system X,Y,Z (the Observing System, OS) 
with a common origin, the Z-axis coinciding with the LOS and the X-axis pointing to the axis crossing the sun center in a direction parallel to the LOS. Note that the axis,
X, SX, Z, and SZ are coplanar and SY and Y are collinear. The real
horizontal velocities, $\bf v$($v,\phi)$, are contained in the (SX,
SY) plane whereas we observe their projections {\bf $\bf
  v'$}($v',\phi')$, in the (X,Y) plane. Calculations based on
Figure~\ref{F:1} lead to the following relationships (equations
\ref{eq:1} and \ref{eq:2}) between the magnitudes $v$ and $v' $, and
the azimuths $\phi$ and $\phi' $, that will allow us to deproject our
observations.

\begin{equation}
v'~^2=v^2 (\sin^2\phi + \cos^2\phi ~ \cos^2 \theta)
\label{eq:1}
\end{equation}

\begin{equation}
\tan \phi'= \frac{\tan \phi}{\cos \theta}
\label{eq:2}
\end{equation}

Since the X-axis and Y-axis do not necessarily coincide with our
natural reference system (i.e the edges of the CCD frame), the
measurement of the projected azimuths, $\phi '$, requires the
knowledge of the orientation of our FOV with respect to the X-axis 
that points to the solar disc center.

\subsection{Flowmaps for different penumbral configurations}
\label{sec:flowmaps}

\begin{figure}
\begin{tabular}{l}
\includegraphics[width=1.\linewidth]{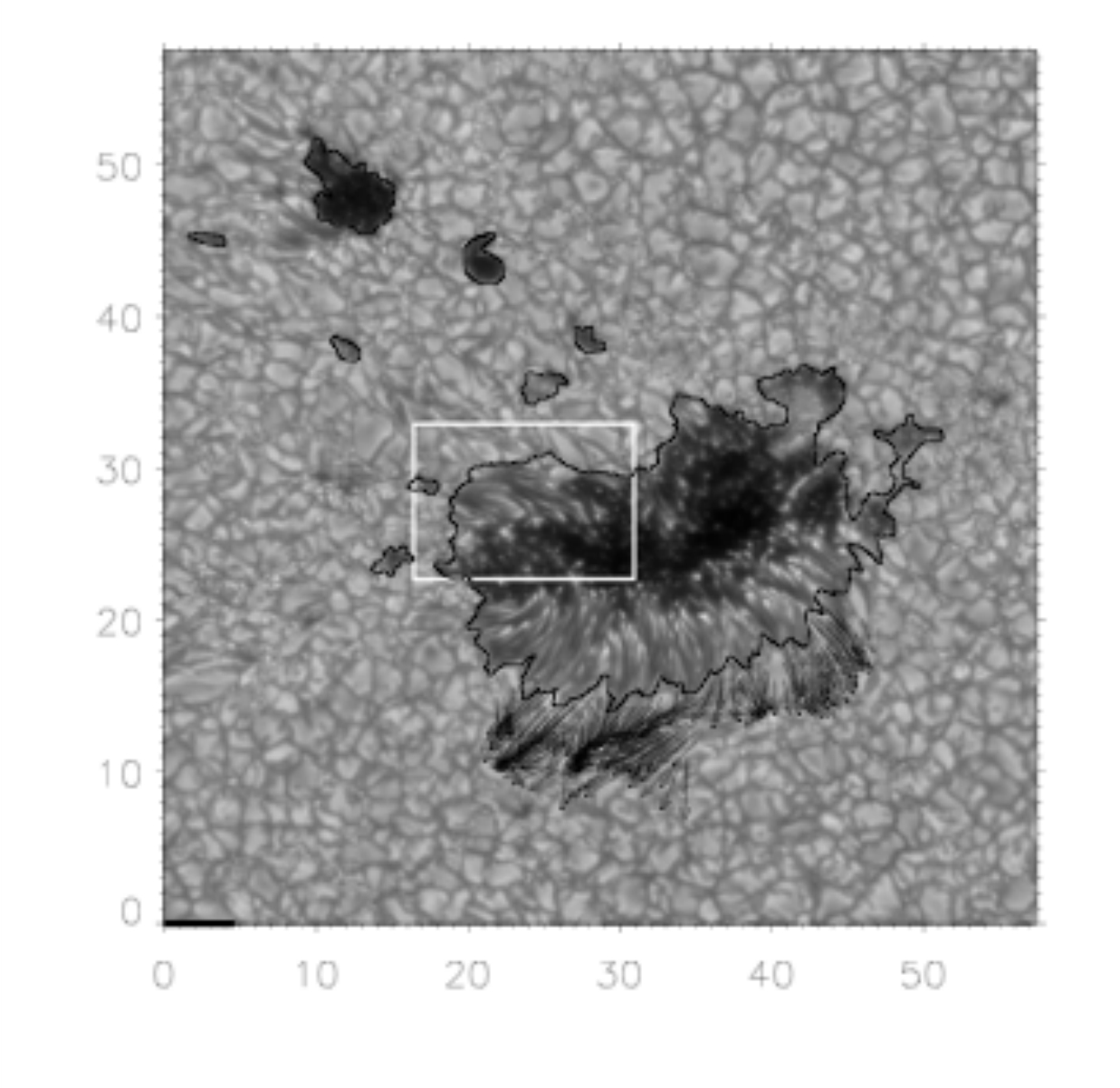}
\end{tabular}
\caption[]{Sunspot {\bf S1}: Map of the horizontal velocities inside the moat with de-projected magnitudes $>$ 0.3~km\,s$^{-1}$. The sunspot exhibits a well developed penumbra in the lower region. The moat flow is clearly found in the same region following the penumbra filamentary direction and in no other region surrounding the sunspot. The coordinates are expressed in arc sec. The black bar at (0,0) represents 1.5 km\,s$^{-1}$ for the projected velocities in all the flow maps hereafter. A peculiar region is white squared.}
\label{F:2}
\end{figure}

\begin{figure}
\centering
\begin{tabular}{l}
\includegraphics[width=1.\linewidth]{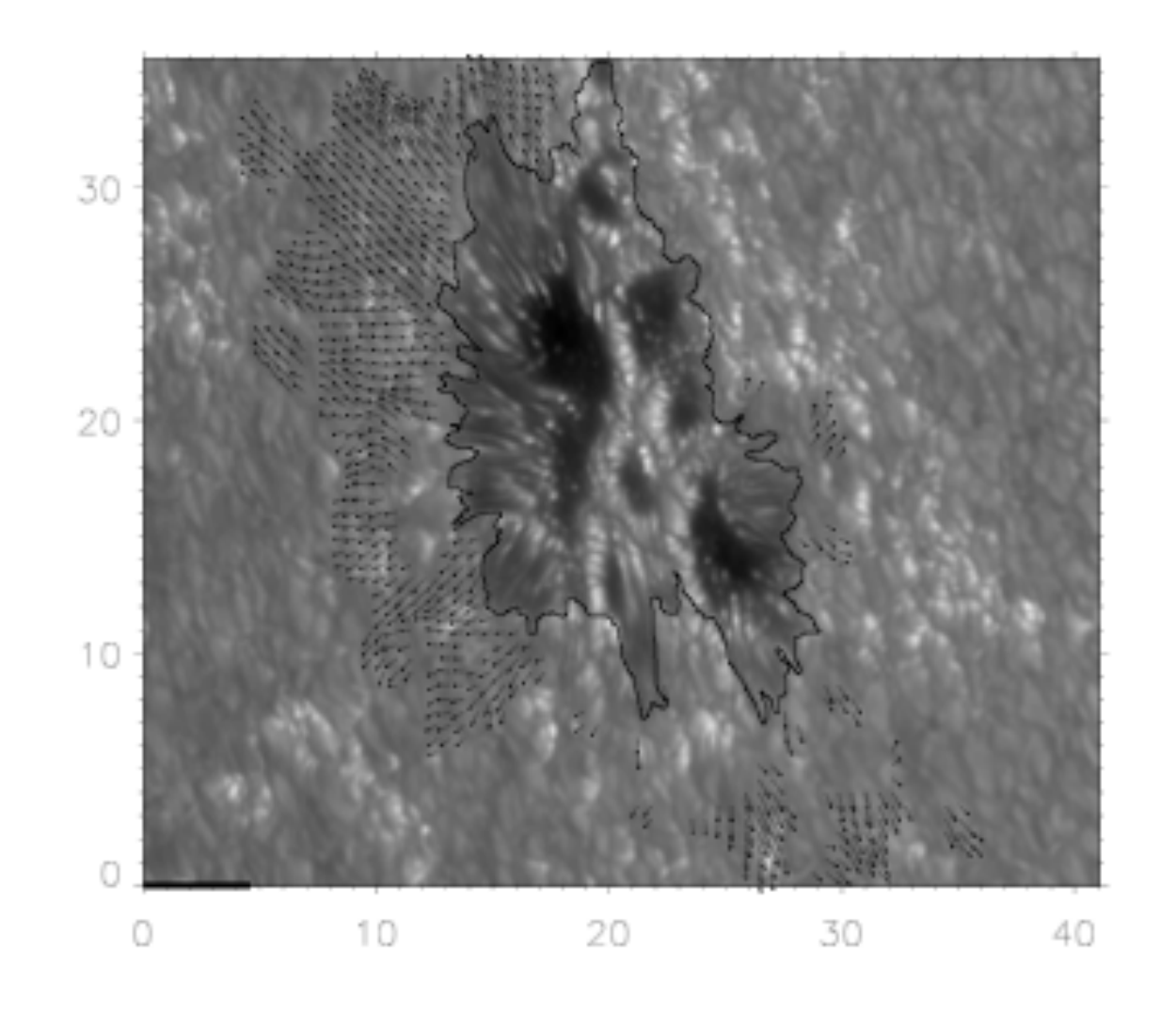}
\end{tabular}
\caption[]{Sunspot {\bf S2}: Map of the horizontal velocities inside the moat with de-projected magnitudes $>$ 0.3~km\,s$^{-1}$. This sunspot shows different penumbrae shapes all around it, from short to long and from wide to narrow penumbrae. Except for the places where the umbra is adjacent to the surrounding photosphere, we identified moat flows all around the sunspot in roughly the same configuration of the penumbra. The coordinates are expressed in arc sec.}
\label{F:3}
\end{figure}

\begin{figure}
\centering
\begin{tabular}{l}
\includegraphics[width=1.\linewidth]{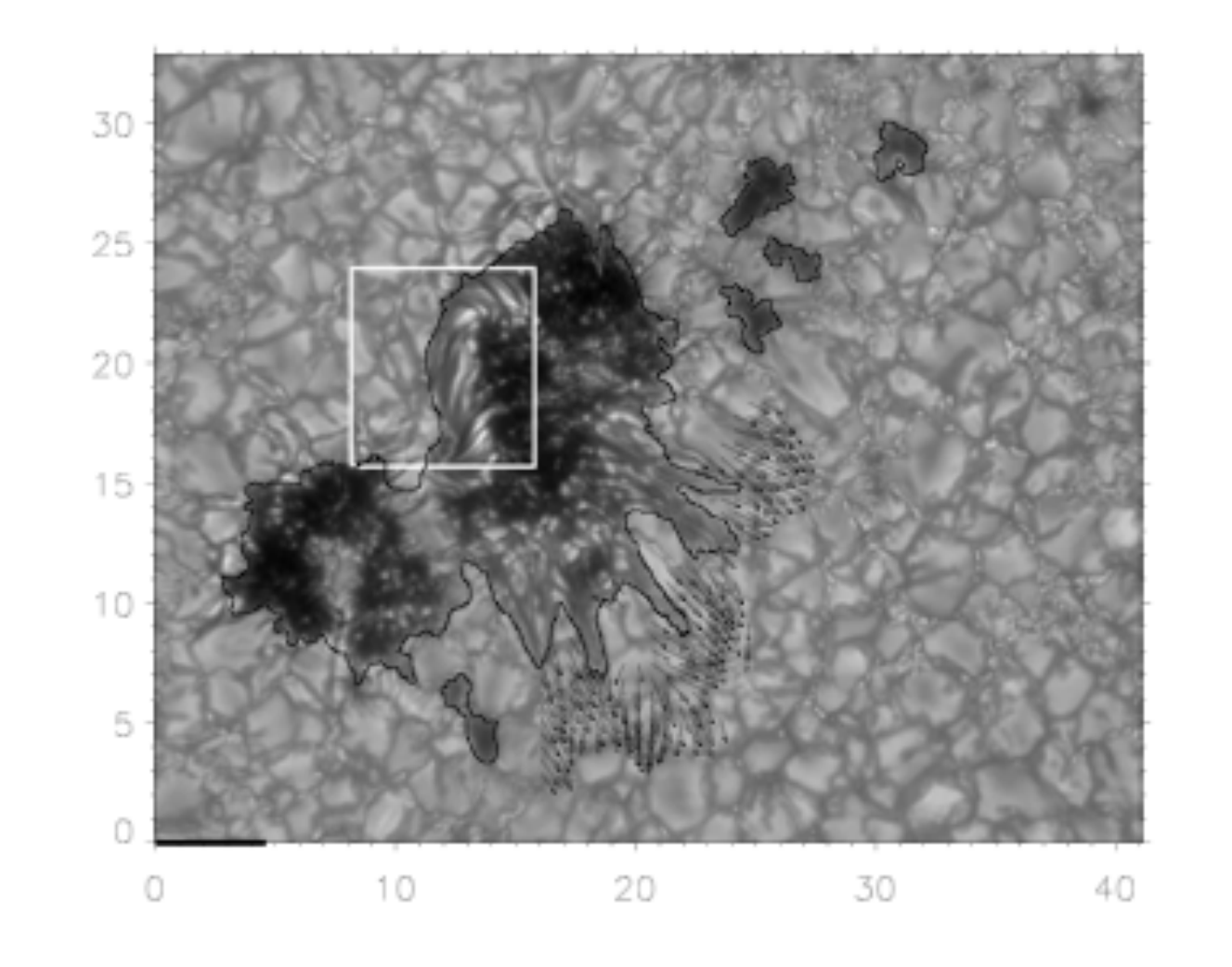}
\end{tabular}
\caption[]{Sunspot {\bf S4}: Map of the horizontal velocities inside the moat with de-projected magnitudes $>$ 0.3~km\,s$^{-1}$. The figure corresponds to a sunspot with two distinct penumbrae. One tangential to the umbral core and the other one coming out radially from the right part of the umbra. The moat flow continues as prolongations of the last one.  The coordinates are expressed in arc sec. A peculiar region is white squared.}
\label{F:4}
\end{figure}

\begin{figure}
\centering
\begin{tabular}{l}
\includegraphics[width=1.\linewidth]{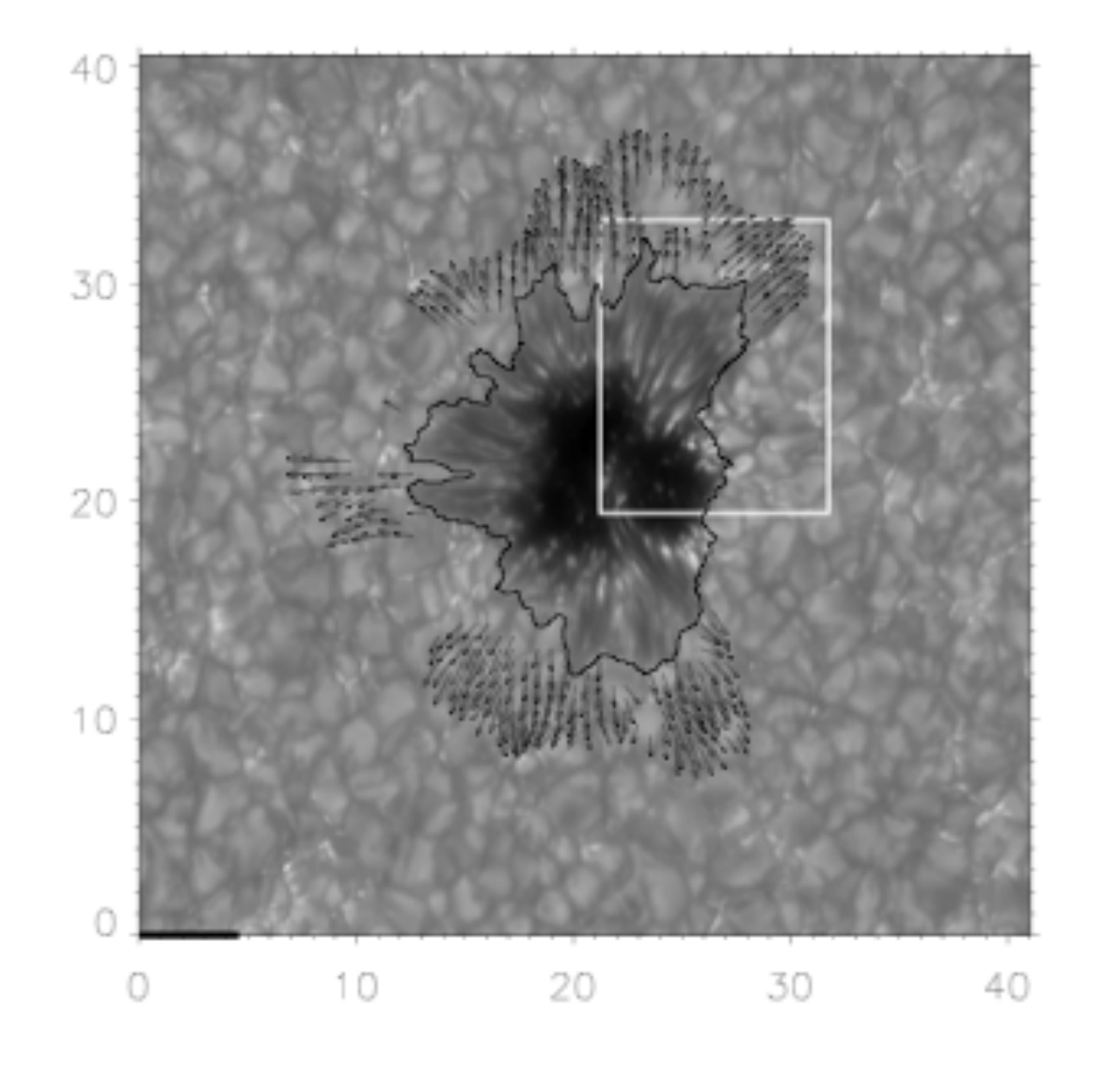}
\end{tabular}
\caption[]{Sunspot {\bf S5}: Map of the horizontal velocities inside the moat with de-projected magnitudes $>$ 0.3~km\,s$^{-1}$. This sunspot presents very distinct areas with and without penumbrae. The penumbral distributions follow radial directions coming out from the umbra. 
Moat flows are organized radially as the penumbra. The coordinates are expressed in arc sec. A peculiar region is white squared.}
\label{F:5}
\end{figure}

\begin{figure}
\centering
\begin{tabular}{l}
\includegraphics[width=1.\linewidth]{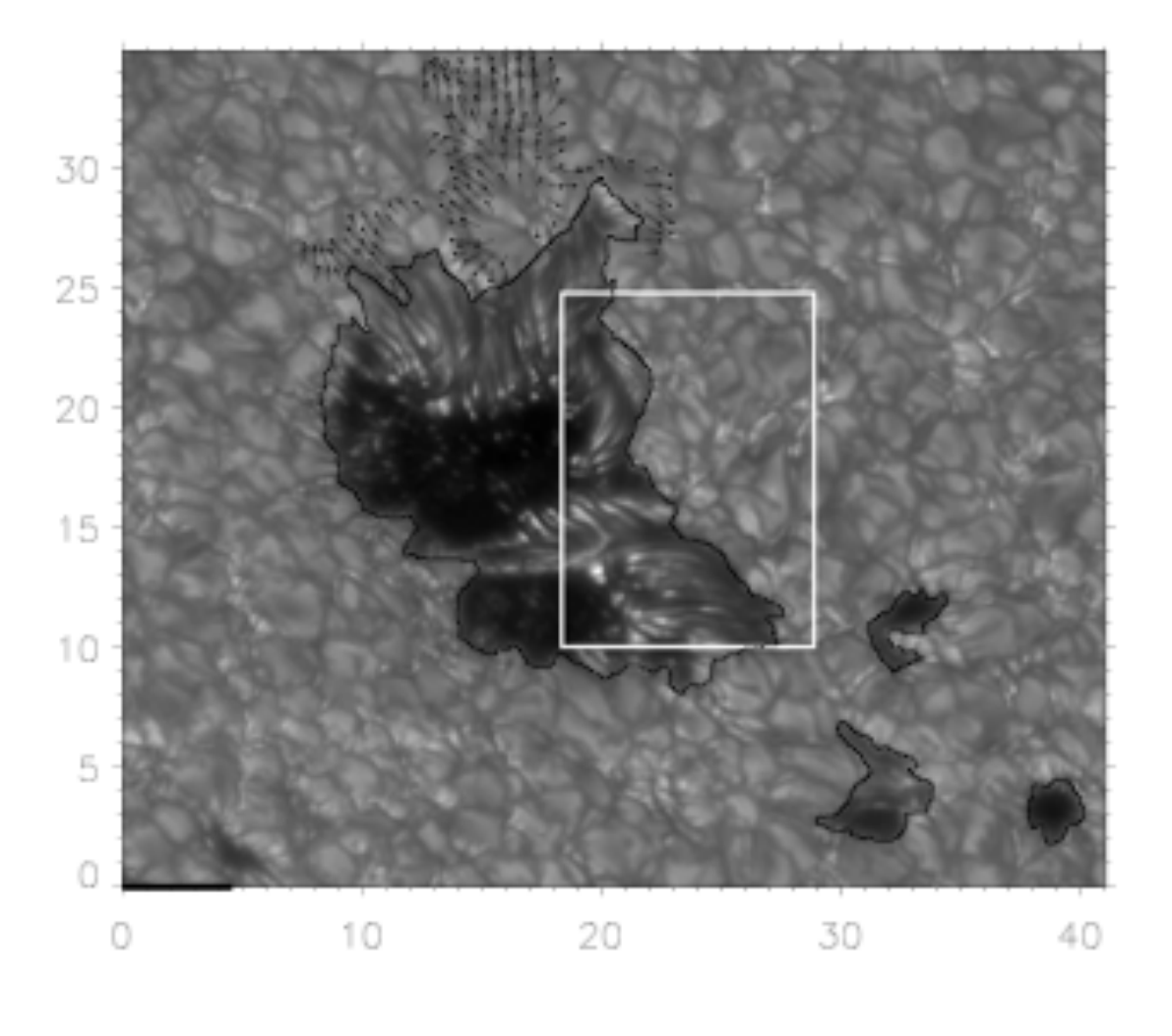}
\end{tabular}
\caption[]{Sunspot {\bf S6}: Map of the horizontal velocities inside the moat with de-projected magnitudes $>$ 0.3~km\,s$^{-1}$.  This sunspot exhibits penumbrae on the right hand and upper sides. The penumbra on the right is mostly tangential to the sunspot border. In the upper penumbra we identified the moat flow as the prolongation of the penumbral filaments as expected. The coordinates are expressed in arc sec.  A peculiar region is white squared.}
\label{F:6}
\end{figure}

\begin{figure}
\centering
\begin{tabular}{l}
\includegraphics[width=1.\linewidth]{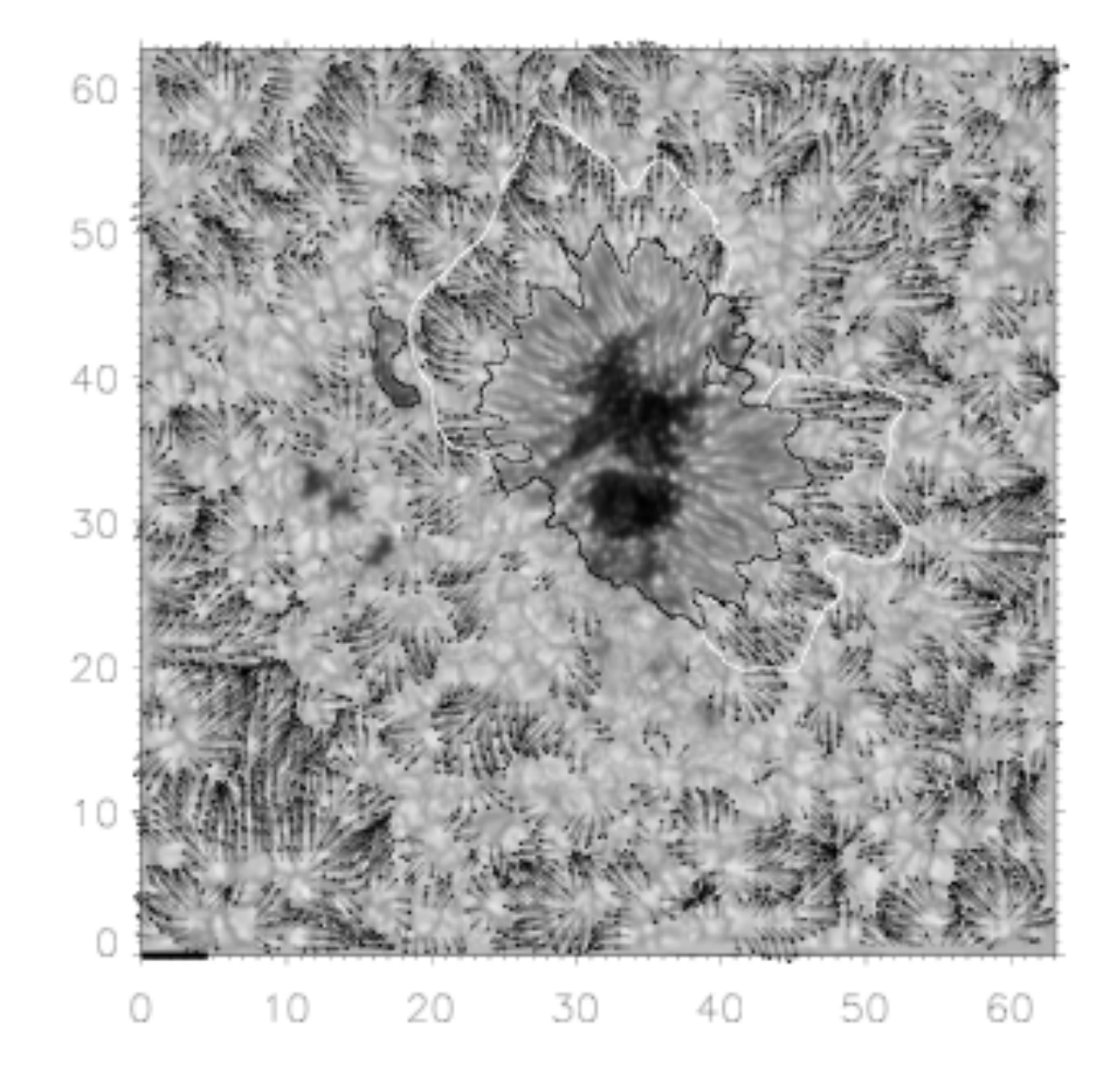} \\
\end{tabular}
\caption[]{Sunspot {\bf S7}: Map of the horizontal velocities surrounding the sunspot with de-projected magnitudes $>$ 0.3~km\,s$^{-1}$. The sunspot includes a well developed penumbra but also an empty region with no penumbra. Looking at the large flows, they completely dissapear in the emptied areas with no prenumbra. The white contours outline the moat regions.  The coordinates are expressed in arc sec.}
\label{F:7}
\end{figure}

As mentioned above, for every sunspot series, the map of horizontal velocities averaged over the whole time series is evaluated. In order to coherently detect the moats and compare the statistics (\S~\ref{sec:statist}) in all active regions of our sample, we have de-projected the observed velocity vectors onto the sunspot plane as described above.
 Overlaying the observed images, we construct maps showing in the granulation field only those projected velocity vectors having de-projected magnitudes above a certain threshold. In these maps, extensive organized outflows coming out from the sunspots can easily be identified so that we can outline masks delimiting in a qualitative way the area of the moats for the purpose of statistical calculations and graphical representation.
We have fixed the mentioned threshold to $\sim$~0.3~km\,s$^{-1}$ as the best compromise to define the limits of the moats. This value should not be understood as an absolute (universal) velocity threshold to define moats in general. As we have mentioned before the particular method used here to measure the displacements may underestimate the velocity magnitudes. Moreover, outside the moat one can easily find velocities larger than that. The point is that this particular threshold makes in our case more evident the existence of a well organized radial outflows around the sunspots in comparison with the rest of the FOV.  Lower values (even zero) for the threshold do not extend the areas of organized flows around the sunspots but produce maps with a very dense and noisy (exploding granules everywhere) representation of arrows where the outline of the frontiers of the moats becomes more difficult. Figures~\ref{F:2} to \ref{F:6} show within the moats, only those velocity vectors with de-projected magnitudes above 0.3~km\,s$^{-1}$. In Figure~\ref{F:7} we extend this representation to the entire granulation field showing that large velocities are also present outside the moat. These velocities are generally grouped and associated with exploding granules.
Our sunspot sample includes different penumbral configurations as a key factor to establish the moat-penumbra relation in a robust way.

Firstly we focus on granulation regions that display moat flows. Close inspection of Figures~\ref{F:2} - \ref{F:7} , and of the upper panel of Figure \ref{F:9}  reveals that the velocity vectors in the moats are oriented following the direction of the penumbral filaments. {\it We observe that in all cases the moat flow direction lines up with penumbral filaments that are oriented radially with respect to the sunspot center.} Nevertheless, the completeness of the velocity vectors (density of arrows) in the flow maps depends on the threshold previously imposed, as expected. This can be seen in Figure~\ref{F:5} where the left-hand side penumbra does not seem to be strictly associated with large flows in the
granulation region. However, this penumbral part has in fact an
associated moat flow similar to the other penumbral regions which is
not visible in the representation since the magnitude of the
velocities is slightly lower than the threshold of
0.3~km\,s$^{-1}$. 
 
Next we consider granulation regions close to the sunspots that do not display moat flows. All sunspots of our sample, except S3, have parts of umbral core in direct contact with granulation regions without intervening penumbrae. In all these granulation regions we do not detect systematic large-scale outflows that fulfill our criteria for moat flows. {\it This supports the conclusion that umbral core boundaries with no penumbra do not display moat flows.}

\begin{figure}
\centering
\begin{tabular}{l}
\includegraphics[width=0.7\linewidth]{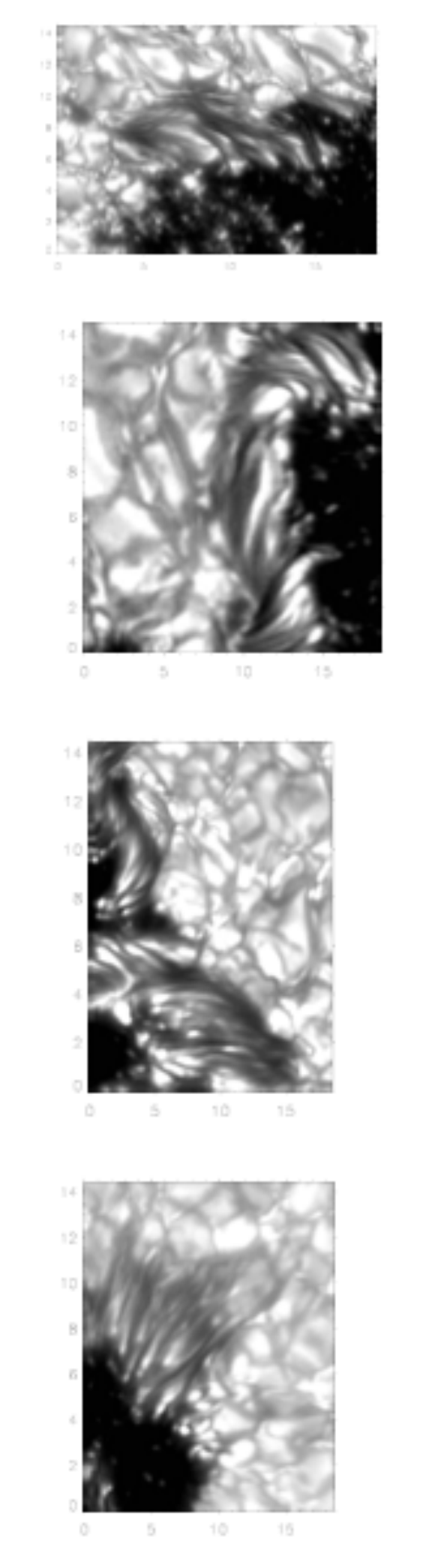} 
\end{tabular}
\caption[]{From top to bottom: Close-ups for peculiar regions in sunspots S1, S4, S6 and S5 respectively. The upper three regions show a penumbra extending tangential to the sunspot border. The lower pannel shows a radial penumbra. The coordinates are expressed in arc sec.}
\label{F:8}
\end{figure}

Finally, there are also granulation regions in the vicinity of penumbrae that lack significant moat flows. To study these cases in more detail we mark in Figures~\ref{F:2}, \ref{F:4}, \ref{F:5} and \ref{F:6} peculiar regions of interest with rectangular white boxes. These regions are represented as close-ups in Figure~\ref{F:8}. In the three upper panels, the penumbral filaments display significant curvature such that they are not radially oriented with respect to the sunspot centre but they extend in a direction tangential to the sunspot border (as marked with black lines in the overview images Figures~\ref{F:2}, \ref{F:4} and \ref{F:6}). 
The lower panel in Figure~\ref{F:8}  shows a penumbra extending radially from the umbra. The surrounding photosphere exhibits moat flows only in the direction of the penumbral filaments (also see Figure~\ref{F:5}). {\it These four examples lead to the conclusion that moat flows are not found in directions transverse to the penumbral filaments.}

In a few cases we found small pores in the vicinity of penumbrae in a
region where we would expect moat flows. Such cases are seen around
coordinates (19,29) in Figure~\ref{F:2} and coordinates (29,11) in
Figure~\ref{F:6}. For these regions the measured proper motions are less reliable since we have a FWHM=$1\farcs0$ tracking window acting on a region of only a
few arc seconds ($\sim$ 2-3) between the pore and penumbra. Another possibility is that the pores themselves are somehow blocking the large outflows changing the expected behavior. In the case presented in Figure~\ref{F:6} the most plausible explanation for the absence of moats is the presence of a neutral line close to coordinates (29,11) as we shall describe in \S~\ref{sec:neutral}.

The findings described above suggest a link between the moat flows and the Evershed flows in penumbrae.  We come back to this relation in  \S~\ref{sec:dis}.

\subsection{Neutral lines affecting the flows behavior}
\label{sec:neutral}
\begin{figure}
\centering
\begin{tabular}{l}
\includegraphics[width=1.\linewidth]{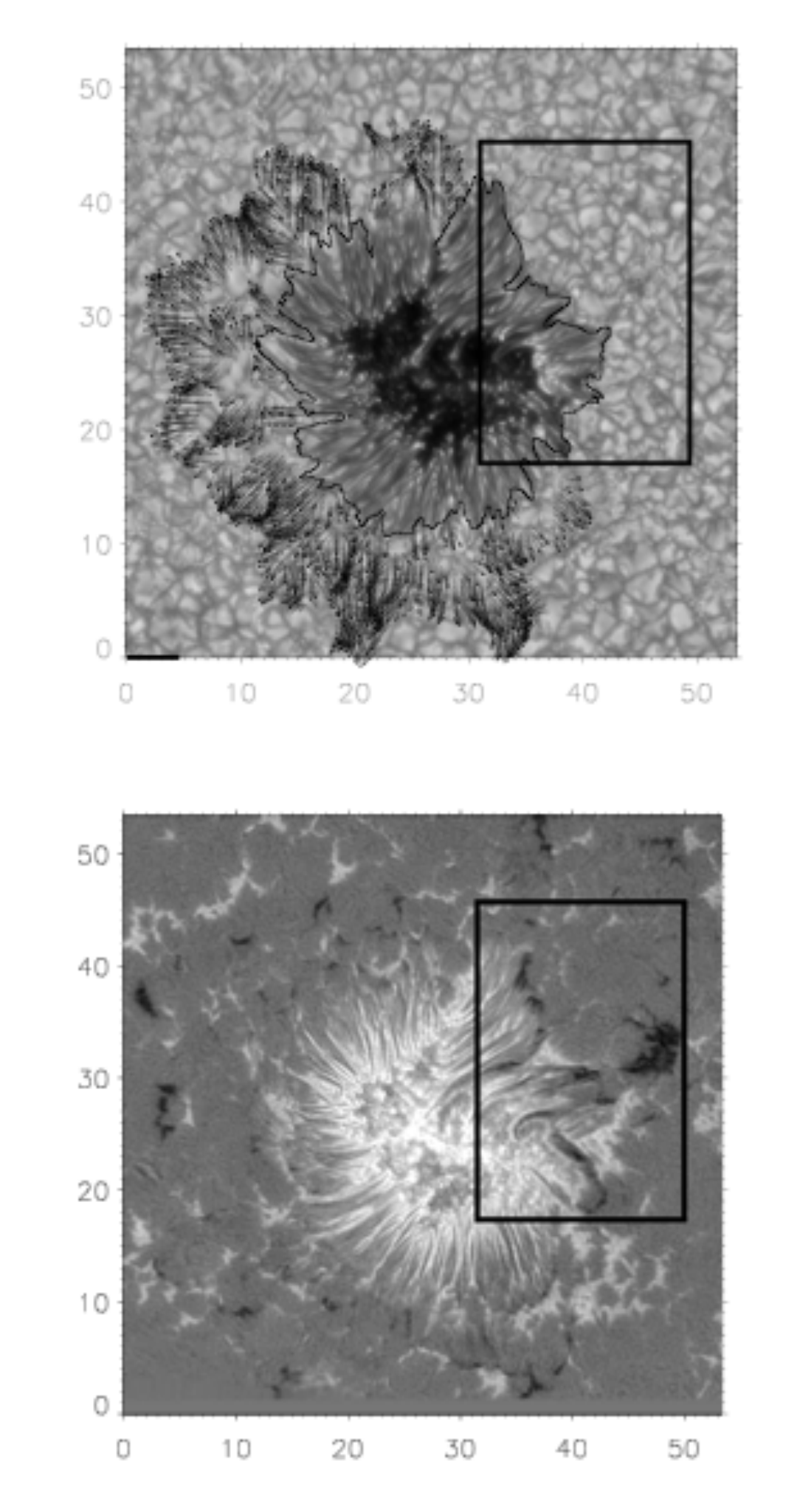} 
\end{tabular}
\caption[]{Sunspot {\bf S3}: {\it Upper panel}: Map of the horizontal velocities with deprojected magnitudes $>$ 0.3~km\,s$^{-1}$. The sunspot in the figure is surrounded by a well-developed penumbra. Large outflows are not found in part of the right-hand side of the spot. When looking at the magnetogram for the same field-of-view ({\it lower panel}) we found a neutral line which is  suspiciously acting as a blocking agent for the flow coming out from the spot. The black box is a common region for both figures. The coordinates are expressed in arc sec. The black bar at (0,0) in the flow map represents  1.5 km\,s$^{-1}$ for the projected velocities.}
\label{F:9}
\end{figure}

Only one of the sunspots we study displays a complete regular and well-developed penumbra surrounding completely the umbral core. Figure~\ref{F:9} shows the flow map calculated for this active region, plotting the horizontal velocities surrounding the sunspot with deprojected magnitudes $>$ 0.3~km\,s$^{-1}$. Large outflows are not found in a certain part of the right-hand side of the spot in the black squared region in the upper panel of Figure~\ref{F:9}.

Following the findings of \S~\ref{sec:flowmaps}, we would expect to
find a moat flow in this region. The penumbral filaments are oriented
radially from the umbral core which suggest the presence of a moat
flow in the granulation region in the direct vicinity. Even with a
lower velocity threshold, no moat flow can be discerned in this
region. Nevertheless, when comparing with the magnetogram (see lower panel of
Figure~\ref{F:9}), we found an inversion in magnetic polarity just
outside the right border of the sunspot: the magnetogram displays
positive polarity (in white) for the sunspot but negative polarity
for the small magnetic elements and pore just outside the
penumbra. The reversal of polarity (or neutral line) is confined to
a narrow region that roughly coincides with the sunspot border. The absence of large outflows following this penumbra is suggestively related to the presence of this neutral line which might somehow be acting as a blocking agent for the moat flow in this region. The position of the neutral line measured by LOS magnetograms is generally influenced by the location of the sunspot ($\theta$ angle). In this case since the sunspot is very close to the disc center we can claim a reliable determination of the neutral line.

For the sunspot in Figure~\ref{F:6} there is also a neutral line crossing all along the right-hand border of the sunspot from top to bottom around coordinates (20,28) to (28,11) respectively (see Figure~2 in \cite{depontieu2007}). The penumbral filaments extend from the umbra to the right and seem to bend, being forced to follow the direction of the neutral line as they approach it.  The sheared configuration of this penumbra is arranged so that the penumbral filaments end up along directions parallel to the sunspot border. Moat flows are then not found beyond this sort of penumbral configuration as mentioned in \S~\ref{sec:flowmaps}.
A similar case was also found by \citet{vargas2007} in a complex $\delta$-configuration active region. For that active region the authors found a strong sheared neutral line crossing a penumbral border where a moat flow was expected to follow the penumbral filaments direction but actually not detected. 
More observations of complex active regions with neutral lines present
in the vicinity of penumbrae are needed to firmly establish
the relation between the absence of moat flows and magnetic neutral
lines.

\subsection{Statistics of velocity fields in moats vs. quiet granulation}
\label{sec:statist}

In this section the statistics of time-averaged horizontal velocity fields in moats is performed. For the sake of comparison, the velocity statistics in quiet granulation areas has also been computed. To that aim, boxes  ($\sim 9\times9$ $arcsec^2$) in regions of less-magnetized (quiet) granulation and far from the moat flows have been manually selected in every FOV. Table~\ref{table3} summarizes the statistical properties of the de-projected magnitudes of horizontal velocity fields within both moat masks and selected granular boxes. Here we again remark that because of the possible underestimation of the displacements by using LCT, we intend to perform not an absolute but a comparative statistical study between moat regions, where we find radially organized outflows, and quiet granulation regions far from the sunspots.

For each sunspot of our sample three velocity magnitude maps result from averaging over different time periods, namely: 5 min, 10 min and the whole duration of the corresponding time series \footnote{Actually, when averaging over 5 min (or 10 min) we obtain more than one map since we then compute maps every 5 min (or 10 min) up to complete the total duration of the series (i.e. when having 40 min of duration for a time series, we would finally obtain 8 maps of 5 min or 4 maps of 10 min). This enables us to improve the statistics of the velocity magnitudes and reduce the noise in our calculations.}.
Table~\ref{table3} shows that in most cases the mean velocity magnitude in moats is greater than in granulation. In a few cases and only for averages in short time periods (5, 10 min), we find similar values in both or even slightly lower velocities in moats. However, when averaging over a long time period, the difference of mean velocities in moats and granulation is in all cases positive and more conspicuous than for short time period averages (see column 10 in the Table). A similar behavior is obtained for the rms parameter (see column 9 in the Table). The maximum velocity values are also systematically larger in moats than in granulation. 
The described statistical behaviour is expected since for short time averaging periods, of the order of the granulation lifetime (5-16 min; see \citet{hirzberger1999} and references there in), the proper motions of the granulation structures compete in magnitude with the velocity of large-scale flows. However, averaging velocity components over long time periods results in local velocity cancellations in short-lived structures while steady motions at large spatial scales prevail. So moats can be considered as a long-term and large-scale phenomenon where the mean velocity exceeds that of quiet granulation by about 30\%.

The significant fluctuations in $\Delta$(rms) and $\Delta$(mean)  (columns 9 and 10 in Table~\ref{table3}) for the velocity fields averaged over the whole time series, deserve a comment and possibly future improved measurements. They could be ascribed to morphological differences in the various cases considered or they could be related to the particular evolutionary state or magnetic field strength in the various sunspots of our sample. Furthermore, the accuracy in the measurement of $\theta$ and $\phi '$ in equations (\ref{eq:1}) and (\ref{eq:2}) is crucial for the determination of $v$. Thus, two sources of inaccuracy are: a) the assumption of constant heliocentric angle $\theta$ all along the sunspot area (this has important impact in regions far from the solar disc centre); and b) the determination in our FOV of the direction pointing to the solar disc centre which defines the X-axis of the Observing Coordinate System, i.e. the angular origin for $\phi '$. 

\begin{figure*}
\centering
\begin{tabular}{l}
\includegraphics[width=.9\linewidth]{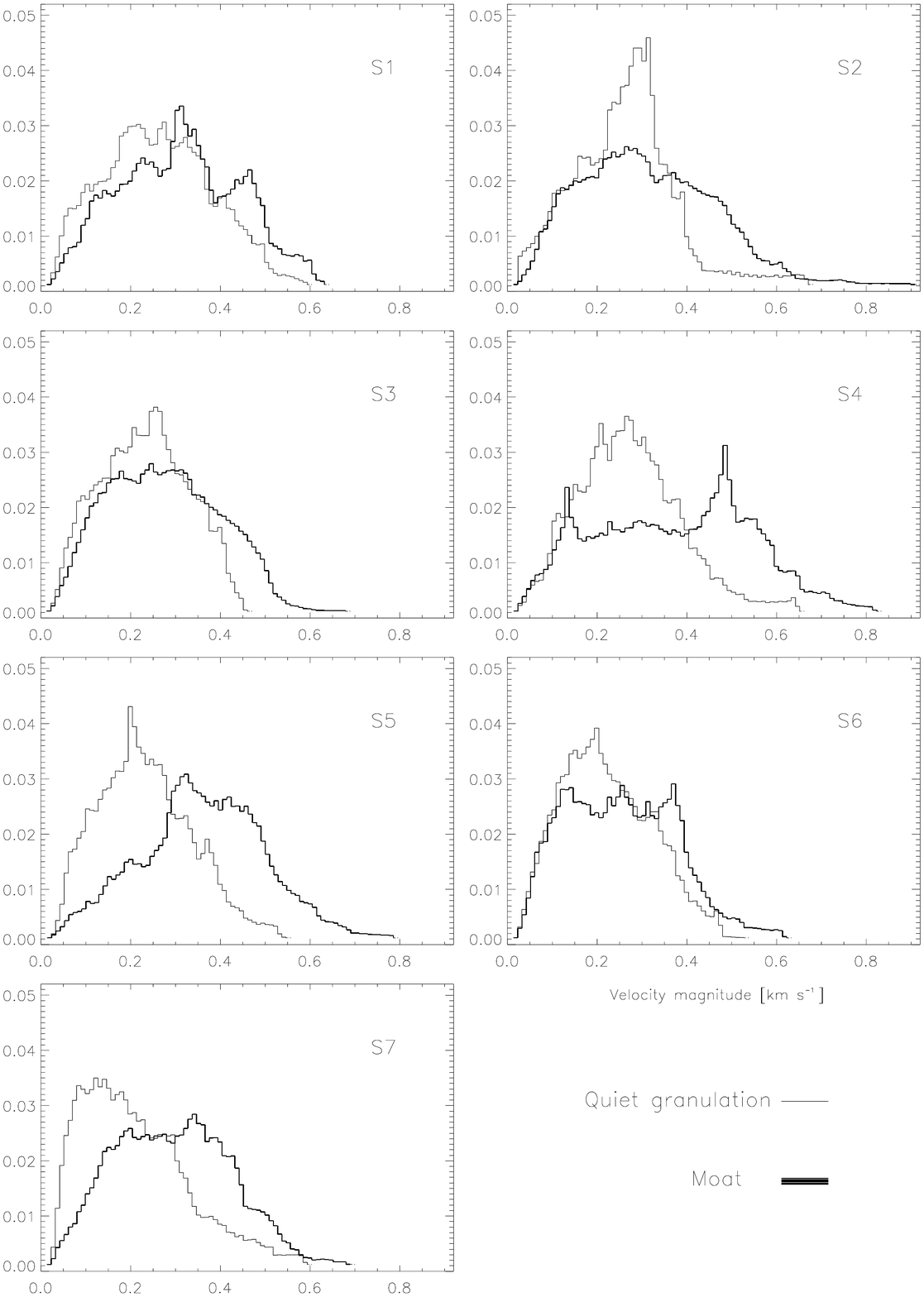} \\
\end{tabular}
\caption[]{Histograms of the velocity magnitudes in moats (\emph{thick line}) and quiet granulation (\emph{thin line}), averaging over the whole time series (more than 40 minutes in every case) for sunspots S1 to S7.  The vertical and horizontal axis represent the percentages and the velocity magnitudes [km\,s$^{-1}$] respectively.}
\label{F:10}
\end{figure*}

Figure~\ref{F:10} shows the histograms of de-projected velocity magnitudes for averages over the whole time series for sunspots S1 to S7. Thick lines correspond to velocities inside the moats and the thin ones to velocities in quiet granulation boxes. In all cases we find the same general trend in both histograms. The thick histograms are globally shifted toward the right with respect to the thin ones.
The left wings of the histograms for granulation lie systematically above those of the histograms for moats, and for S1, S2, S3, S4 and S6, both histograms nearly coincide at the most left end. This means that within the moats standard velocity values for quiet granulation are still present but with a lower weight. Such is the case for the fragments of exploding granules that additionally are swept by large-scale moat flows. 
At some point from left to right, both histograms cross each other so that the right wing of the moat histogram surpasses the corresponding quiet granulation wing and extends to larger velocity values ($>$ 0.6~km\,s$^{-1}$). This confirms the predominance of large velocities in moats. The intersection point of both histograms corresponds to  $\sim$ 0.3~km\,s$^{-1}$ in almost all cases (a bit higher in S1, S2 and S4). 
Interestingly, this value coincides with the threshold empirically selected a priori ( \S~\ref{sec:flowmaps}) to clearly distinguish the frontiers of the moats, which supports the goodness of the limit-value (0.3~km\,s$^{-1}$) chosen to detect moats in our data set. 

\begin{table*}[]
\caption{Statistics of de-projected horizontal velocity magnitudes [$m$ $s^{-1}$]}
\centering
\begin{tabular}{|c|c|r|r|r|r|r|r|c|c|}
\hline
 Sunspot & Duration & \multicolumn{3}{c|}{Moat} & \multicolumn{3}{c|}{Granulation} &  $\Delta$(rms) \footnotemark[\dag] &  $\Delta$(mean) \footnotemark[\dag]    \\
 & (min) & max & rms & mean & max & rms & mean & \% & \%  \\ \hline
 
\multirow{3}{*}{S1} & 5 & 1048 & 174 & 333 & 947 & 157 & 315 & 10.8 & 5.7 \\
 & 10 & 947 & 162 & 315 & 842 & 157 & 314 & 3.2 & 0.3 \\
 & 53 & 639 & 137 & 301 & 596 & 123 & 250 & 11.4 & 20.4\\ 
 \hline
\multirow{3}{*}{S2} & 5 & 1187 & 183 & 329  & 898 & 162 & 307 & 13.0 & 7.1\\
 & 10 & 1278 & 182 & 327  &  862 & 155 & 300 & 17.4 & 9.0 \\
 & 47 & 921 & 150 & 303 & 678 & 115 & 249 & 30.4 & 21.7
\\ \hline
\multirow{3}{*}{S3} & 5 & 1036 & 146 & 290 & 870 &  147 & 282 & -0.7 & 2.8\\
 & 10 &  868 & 146 & 281 & 671 & 137 &  289 & 6.6 & -2.8 \\
 & 47 & 689 & 126 & 265 & 451 & 101 &  220 & 24.8 & 20.5
\\ \hline
\multirow{3}{*}{S4} & 5 & 1060 & 198 &  411 & 863 & 164 & 362 & 20.7 & 13.5 \\
 & 10 & 954 & 180 & 379 & 780 &  154 &  348 & 16.9 & 8.9\\
 & 45 & 834 & 179 & 365 &  655 &  118 &  262 & 51.7 & 39.3
 \\ \hline
\multirow{3}{*}{S5} & 5 & 999 & 178 & 355 & 968 & 172 & 307 & 3.5 & 15.6\\
 & 10 & 972 & 187 & 372 & 1017 & 159 & 291 & 17.6 & 27.8 \\
 & 40 & 802 & 141 & 355 & 545 & 107 & 220 & 31.8 & 61.4
 \\ \hline
\multirow{3}{*}{S6} & 5 & 699 & 129 & 257 &  683 &  130 &  278  & -0.8 & -7.6\\
& 10 & 673 & 131 & 248 & 723 & 136 & 277 & -3.7 & -10.5 \\
& 40 & 622 & 125 & 243 &  526 & 104 & 210 & 20.2 & 15.7\\ 
 \hline
\multirow{3}{*}{S7} & 5 & 872 & 158 & 300 & 759 & 142 & 226 & 11.3 & 32.7 \\
& 10 & 785 & 159 & 320 & 828 &  149 & 227 & 6.7 & 41.0\\
& 40 & 692 & 131 & 285 & 593 & 121 & 196 & 8.3 & 45.4
  \\ \hline 
\hline

%
%
\end{tabular}
\footnotetext[\dag]{$\Delta$(rms) and  $\Delta$(mean) stand for increments of the rms and mean values in moats with respect to quiet granulation. }
\label{table3}
\end{table*}

\section{Discussion}
\label{sec:dis}

Seven time series of sunspots were restored from instrumental and
atmospheric aberrations using MFBD and MOMFBD techniques. The high
quality of the images allowed us to study proper motions of granules
outside the sunspots and measure their time-averaged velocities.

We have extended the study by \citet{vargas2007} to a larger sample of active regions and
systematically confirmed their findings: a) moat flows are oriented following the direction of the penumbral filaments; b) in granulation regions found adjacent to an irregular penumbral side parallel to the penumbral filaments moats are absent, or in other words, moats do not develop in the direction transverse to the penumbral filaments.  Note that if the moat flows were originated by the blockage of the heat flux from below by the penumbra, one would expect moat flows directed along, but also, transverse to the direction of the penumbral filaments; c) umbral core sides with no penumbra do not display moat flows. Moreover we include in our sample a case in which a neutral line extends along a penumbral border where we would expect a moat flow continuation. For this sunspot we do not find any moat flow following the direction of the penumbral filaments after crossing the penumbral border where we see a change in magnetic polarity. The same result is found by \citet{vargas2007} in a penumbral portion of a complex active region crossed by a strong sheared neutral line.

All these results indicate a likely connection between the moat flows and flows aligned with penumbral filaments.  In a recent work, \citet{cabrera2006} suggest that the Evershed clouds inside penumbrae propagate to the surrounding moat and then become MMFs after crossing the sunspot border. The MMFs displacements trace very well-defined paths that can actually be clearly seen when averaging magnetograms in time \citep{sainz2005}. Some of these MMFs are seen to start inside the penumbra.

In this paper we also complement the study by \citet{vargas2007}  with a statistical analysis describing the differences between velocity fields in moat flows and in less-magnetized solar granulation nearby. In contrast to the granulation, moat flows are well-organized, steady and large-scale motions. For averages made over more than 40 minutes, the mean velocity in moats (0.3~km\,s$^{-1}$) exceed that of quiet granulation ($\sim$ 0.23~km\,s$^{-1}$) by about 30\%, although we obtain a considerable dispersion in the results. Also the rms of the velocity magnitude is greater in moats by a similar percentage.

The histograms of velocity magnitudes in the moats are broader than those in granulation. The histograms of granulation show conspicuous maxima, most of them in a range from 0.2 to 0.3~km\,s$^{-1}$, whereas the histograms of moats present a flatter top. Systematically, the low velocity wing of the granulation histogram lies above that corresponding to the moat. At some point about  0.3-0.4~km\,s$^{-1}$, both histograms cross each other and the right wing of the moat histogram extends beyond that of the granulation to larger velocity values ($>$ 0.6~km\,s$^{-1}$).

We have studied one case of a sunspot penumbra displaying a neutral line all along a sector of the penumbra (Figure~\ref{F:9}). This neutral line is detected at the penumbral boundary but also emphasized by a large opposite polarity concentration nearby. Interestingly, this penumbral sector shows no moat flow. We interpret this evidence as an indication
of the Evershed flow being forced to go into deeper subphotospheric layers at a faster pace than what is normally though to occur in penumbral regions not associated to neutral lines \citep{westendorp1997}. A similar case, but in a sheared neutral line of a $\delta$-spot, was found by \citet{vargas2007}.

Though there is increasing evidence linking the moat flows and the Evershed flow along the penumbral filaments, the debate regarding the existence of a moat flow around umbral cores and individual pores is still undergoing. In a recent work, \cite{deng2007}, found that the dividing line between  radial inward and outward proper motions in the inner and outer penumbra, respectively, survived the decay phase, suggesting that the moat flow is still detectable after the penumbra disappeared. 

Previous works \citep{sobotka1999, roudier2002}, have measured horizontal proper motions in and around pores and have observed a ringlike structure of positive divergence ("rosettas") around the pores, which is related to a continuous activity of exploding granules. \cite{roudier2002} identified a very clear inflow around pores which corresponds to the penetration of small granules and granular fragments from the photosphere into the pores, pushed by granular motions originated in the divergence centres around them. They conclude that the motions at the periphery of the pore are substantially and continuously influenced by the external plasma flows deposited by the exploding granules. We interpret the dividing line between radial inward and outward motions, found by \cite{deng2007} outside the residual pore, as corresponding to the centres of divergence of the exploding granules around the pore. The outward motions these authors described, which are not in the immediate surroundings of the pore but separated by the annular inward motion would then correspond to the flows coming out from the regular mesh of divergence centers around the pore.

The important questions related to the flows inside and outside sunspots are yet being studied. In the present work we contribute with a new sample of sunspots observed between 2003 and 2006 in six different observing campaigns. The data sets have been selected on basis of the seeing quality (sharpness, homogeneity, duration) and on the availability of suitable targets: the presence of spots with some form of irregular penumbra.
In all of our samples we follow the evolution of the sunspots for more than 40 min. Although this only represents a snapshot in the evolution of the sunspots through all their emerging and decaying processes, our sample includes sunspots in different evolutionary stages and penumbral configurations. 

New facilities such as the recently launched HINODE satellite can provide long time series of active regions with a constant image quality and enough spatial resolution to provide firm confirmation of the evidences found in this paper. The addition of simultaneous Doppler and magnetogram data to the continuum intensity (or G-band) data sets
will enhance our understanding of the link between the Evershed and the moat flows. Needless to say, the study of those stages were penumbrae are just formed or destroyed becomes of extreme importance to validate our findings. Similarly, the results that can be expected in the coming years from local helioseismology, describing the flow patterns
in the deeper layers near sunspots, will prove crucial for the establishment of  a clear link between these two well known flow patterns that has so far been not appreciated.   

\acknowledgments{The Swedish 1-m Solar Telescope is operated on the
  island of La Palma by the Institute of Solar Physics of the Royal
  Swedish Academy of Sciences in the Spanish Observatorio del Roque de
  los Muchachos of the Instituto de Astrof\'isica de Canarias. S. Vargas is thankful to A. Sainz Dalda for comments and discussions. Partial support by the Spanish Ministerio de Educaci\'on y Ciencia through project ESP2003-07735-C04 and financial support by the European
  Commission through the SOLAIRE Network (MTRN-CT-2006-035484) are
  gratefuly \mbox{acknowledged}. This research was supported through grants 146467/420 and 159137/V30 of the Research Council of Norway.}

\end{document}